\documentclass[reprint, superscriptaddress, amsmath,amssymb, aps]{revtex4-2}
\usepackage{graphicx}
\usepackage{braket}
\usepackage{amssymb,amsmath}
\usepackage[T1]{fontenc}
\usepackage[english]{babel}
\usepackage{textgreek} 
\usepackage{color}
\usepackage{braket}

\begin{document}

\title{Extreme ultraviolet wave packet interferometry of the autoionizing HeNe dimer}

\author{Daniel Uhl}
\author{Andreas Wituschek}
\author{Rupert Michiels}
\affiliation{Institute of Physics, University of Freiburg, Hermann-Herder-Str. 3, 79104 Freiburg, Germany}
\author{Florian Trinter}
\affiliation{Institut f\"ur Kernphysik, J. W. Goethe-Universit\"at, Max-von-Laue-Strasse 1, 60438 Frankfurt am Main, Germany}
\affiliation{Molecular Physics, Fritz-Haber-Institut der Max-Planck-Gesellschaft, Faradayweg 4-6, 14195 Berlin, Germany}
\author{Till Jahnke}
\affiliation{Institut f\"ur Kernphysik, J. W. Goethe-Universit\"at, Max-von-Laue-Strasse 1, 60438 Frankfurt am Main, Germany}
\affiliation{European XFEL, Holzkoppel 4, 22869 Schenefeld, Germany}
\author{Enrico Allaria}
\author{Carlo Callegari}
\author{Miltcho Danailov}
\author{Michele Di Fraia}
\author{Oksana Plekan}
\affiliation{Elettra-Sincrotrone Trieste S.C.p.A., 34149 Basovizza (Trieste), Italy.}
\author{Ulrich Bangert}
\author{Katrin Dulitz}
\author{Friedemann Landmesser}
\author{Moritz Michelbach}
\affiliation{Institute of Physics, University of Freiburg, Hermann-Herder-Str. 3, 79104 Freiburg, Germany}
\author{Alberto Simoncig}
\author{Michele Manfredda}
\author{Simone Spampinati}
\author{Giuseppe Penco}
\affiliation{Elettra-Sincrotrone Trieste S.C.p.A., 34149 Basovizza (Trieste), Italy.}
\author{Richard James Squibb}
\author{Raimund Feifel}
\affiliation{Department of Physics, University of Gothenburg, Origov\"agen 6 B, 41296 Gothenburg, Sweden}
\author{Tim Laarmann}
\affiliation{Deutsches Elektronen-Synchrotron DESY, Notkestr. 85, 22607
Hamburg, Germany}
\affiliation{The Hamburg Centre for Ultrafast Imaging CUI, Luruper Chaussee 149, 22761 Hamburg, Germany}
\author{Marcel Mudrich}
\affiliation{Department of Physics and
Astronomy, Aarhus University, Ny Munkegade 120, 8000 Aarhus, Denmark}
\author{Kevin C. Prince}
\affiliation{Elettra-Sincrotrone Trieste S.C.p.A., 34149 Basovizza (Trieste), Italy.}
\author{Giulio Cerullo}
\affiliation{IFN-CNR and Dipartimento di Fisica, Politecnico di Milano, Piazza L. da Vinci 32, 20133 Milano, Italy.}
\author{Luca Giannessi}
\affiliation{Elettra-Sincrotrone Trieste S.C.p.A., 34149 Basovizza (Trieste), Italy.}
\affiliation{Istituto Nazionale di Fisica Nucleare - Laboratori Nazionali di Frascati, Via E. Fermi 40, 00044 Frascati, Roma}
\author{Frank Stienkemeier}
\author{Lukas Bruder}
\email{lukas.bruder@physik.uni-freiburg.de}
\affiliation{Institute of Physics, University of Freiburg, Hermann-Herder-Str. 3, 79104 Freiburg, Germany}

\date{\today}

\begin{abstract}
Femtosecond extreme ultraviolet wave packet interferometry (XUV-WPI) was applied to study resonant inter-atomic Coulombic decay (ICD) in the HeNe dimer. 
The high demands on phase stability and sensitivity for vibronic XUV-WPI of molecular-beam targets are met using an XUV phase-cycling scheme. 
The detected quantum interferences exhibit vibronic dephasing and rephasing signatures along with an ultrafast decoherence assigned to the ICD process. 
A Fourier analysis reveals the molecular absorption spectrum with high resolution. 
The demonstrated experiment shows a promising route for the real-time analysis of ultrafast ICD processes with both high temporal and spectral resolution. 
\end{abstract}

\maketitle


\section{Introduction}
In wave packet interferometry (WPI),  the interference between two optically prepared wavepackets is controlled and mapped using a sequence of ultrashort phase-locked laser pulses\,\cite{scherer_fluorescence-detected_1991}. 
WPI is a key concept in coherent control\,\cite{ohmori_wave-packet_2009}, high-resolution metrology\,\cite{lomsadze_tri-comb_2018} and ultrafast multidimensional spectroscopy\,\cite{tekavec_fluorescence-detected_2007, bruder_coherent_2019}. 
However, corresponding interferometric concepts are very scarce in the extreme ultraviolet (XUV) and X-ray spectral domains, despite many promising theory proposals for such experiments\,\cite{mukamel_multidimensional_2013}. 
The major technical obstacles are (i) the demand for extreme phase stability to perform XUV/X-ray WPI, and (ii) the lack of selective/background-free probes to recover the weak nonlinear signals of interest. 
These ingredients were demonstrated in separate experiments: on the one hand, in XUV and soft X-ray interferometry\,\cite{prince_coherent_2016, jansen_spatially_2016, usenko_attosecond_2017, wituschek_tracking_2020,wituschek_phase_2020,kaneyasu_electron_2021, bellini_phase-locked_2001, cavalieri_ramsey-type_2002, koll_experimental_2022}, on the other hand, in background-free NIR-XUV/X-ray four-wave-mixing schemes\,\cite{bencivenga_four-wave_2015, cao_noncollinear_2016, foglia_first_2018, rouxel_hard_2021}. 
Only recently, the combination of both ingredients was achieved in a single experiment by introducing a phase-cycling concept for XUV pulses\,\cite{wituschek_tracking_2020, wituschek_phase_2020}. 
A recent stability improvement of this scheme even shows promise for extensions to interferometric X-ray experiments\,\cite{uhl_improved_2022}.
These achievements lay the basis for the flexible implementation of various nonlinear spectroscopy concepts in the short-wavelength domain. 

The few XUV interferometry experiments reported so far were restricted to the detection of electronic coherences in isolated atoms\,\cite{wituschek_tracking_2020, prince_coherent_2016, kaneyasu_electron_2021, cavalieri_ramsey-type_2002} and vibrational beatings in molecules\,\cite{okino_direct_2015, tzallas_extreme-ultraviolet_2011}.   
Vibrational WP beatings provide only information about the relative frequencies between the involved vibrational modes which can lead to ambiguities in complex and congested spectra e.g. due to similar vibrational signatures in different electronic states.  
In contrast to that, the electronic beat spectrum correlates the vibronic phase evolution directly to the molecular ground state and can thus be used to disentangle the excited electronic states. 
In this letter, we demonstrate the latter approach and track the complex response of vibronic WPs in autoionizing states of a molecule using XUV interferometry. 

Autoionization has been extensively studied in isolated atoms and is caused by the configuration interaction between the electrons of the atom leading to the famous Fano resonance lineshapes\,\cite{fano_effects_1961}. 
In weakly-bound matter as e.g. van-der-Waals clusters, interatomic Coulombic decay (ICD) offers an additional decay mechanism mediated by interparticle dipolar interactions\,\cite{cederbaum_giant_1997}, which has been extensively studied in recent years\,\cite{jahnke_interatomic_2020}. 
Resonant ICD may occur if the resonantly excited bound state of one atom is energetically embedded in the ionization continuum of another atom in the system. 
This process has been studied in the HeNe heterodimer using frequency-domain spectroscopy\,\cite{trinter_vibrationally_2013}. 
Here, we apply XUV-WPI to track the ultrafast decoherence of electronic WPs caused by the resonant ICD process. 
In contrast to frequency-domain spectroscopy, this approach opens up the possibility to study ICD and related ultrafast conversion processes in real time with high spectral and temporal resolution. 

\begin{figure}
\centering\includegraphics[width=0.9\linewidth]{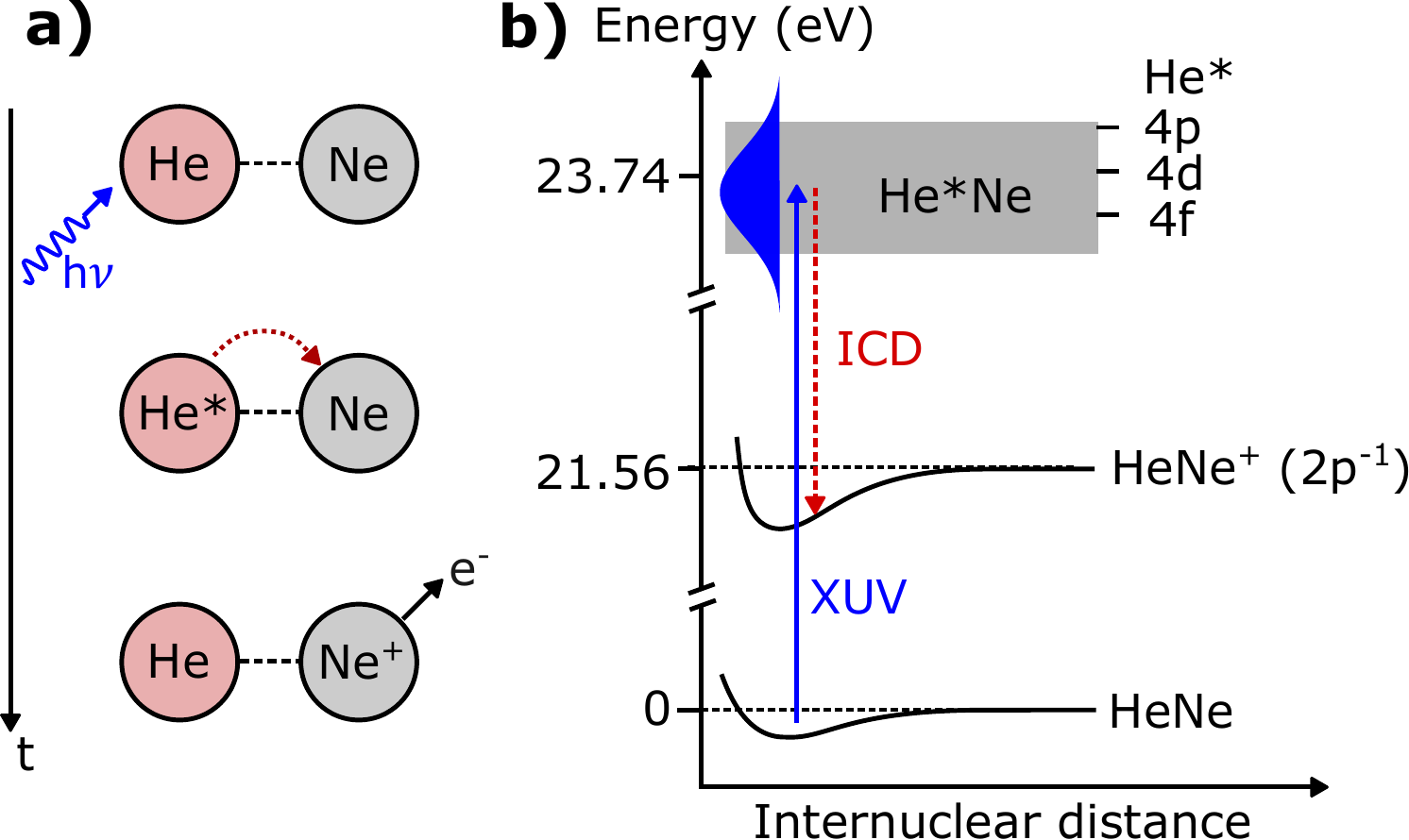}
\caption{Resonant ICD in the HeNe dimer. 
(a) Excitation of the He site with a high-energy photon is followed by an energy transfer to the Ne site, leading to the ejection of a $2p$ electron. 
(b) Sketch of the relevant energy levels of HeNe along with the XUV excitation scheme (blue arrow) and the ICD (red dashed arrow). 
The congested potential energy curves correlating to the $p-, d-$ and $f-$asymptotes are indicated by the grey-shaded area.}
\label{fig1}
\end{figure}
Figure\,\ref{fig1} shows schematically the probed ICD process in the HeNe dimer. 
The two rare-gas atoms form a weakly bound van-der-Waals complex (ground-state binding energy 2\,meV, equilibrium distance 3\,\AA\,\cite{cybulski_ground_1999}). 
The He site is optically excited to the states correlated to the He $n=4$ asymptote (23.7\,eV) lying below the ionization potential (IP) of He (24.59\,eV) but above the IP of Ne (21.56\,eV). 
Subsequent energy transfer between the two sites leads to the ejection of a $2p$ electron from the Ne site. 

\begin{figure}
\centering\includegraphics[width=\linewidth]{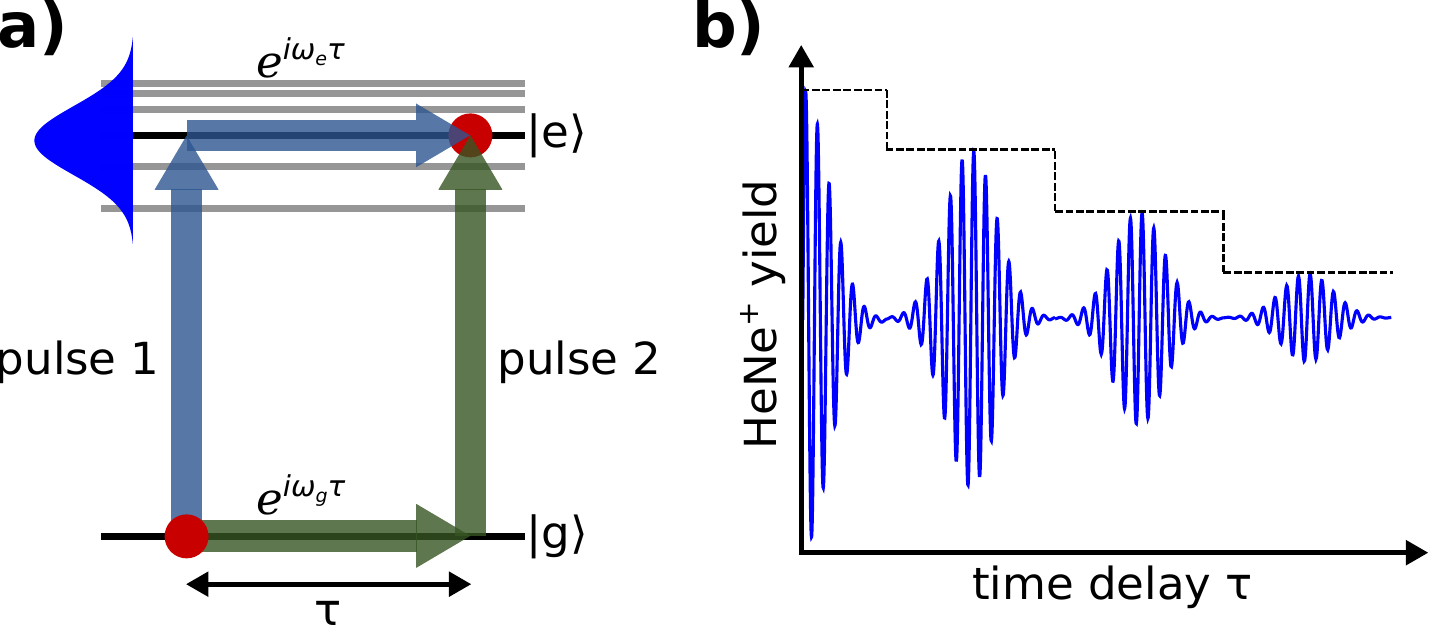}
\caption{WPI scheme. (a) The excitation pathways of pulse 1 and 2 accumulate different phase factors ($\exp [i \omega_j \tau],\, j=g,e$) as a function of the pulse delay $\tau$ which leads to a characteristic interference pattern in the ion yield, as schematically shown in (b). 
The dashed line indicates the expected step-wise decoherence caused by the ICD of the system.}
\label{fig2}
\end{figure}

The process is probed using XUV-WPI. 
The technique is schematically described in Fig.\,\ref{fig2}. 
Each of the two phase-coherent XUV pulses excites a specific pathway in the sample leading to the same final state. 
Since the two pathways evolve along different states during the pulse delay $\tau$, they accumulate different phase factors, giving rise to destructive/constructive interference in the final state. 
The ICD rate reflects the population of the $n=4$ state manifold, and thus maps the pathway interference onto the HeNe$^+$ yield (Fig.\,\ref{fig2}b) which is detected in the experiment. 
The superposition of many pathways excited within the laser bandwidth leads to a complex beat spectrum, which exhibits dephasing and rephasing recurrences and an overall decay due to the decoherence of the system  (Fig.\,\ref{fig2}b). 
In addition, the Fourier transform of the interference fringes provides the excitation spectrum of the sample\,\cite{scherer_fluorescence-detected_1991} with a frequency resolution that can exceed the laser bandwidth by orders of magnitude\,\cite{bruder_phase-modulated_2015}. 
This approach enables a flexible adaptation to the system's time and frequency scales.

Vibronic interferometry requires the control of the relative phase/delay of the excitation pulses to a fraction of the fringe period ($hc/\lambda_\mathrm{XUV}$) which is particularly challenging to achieve at XUV wavelengths, for which the fringe period is on the order of $100-200$\,as. 
This is in contrast to the study of pure vibrational WP beatings, where the fringe periods are typically two to three orders of magnitude longer and thus demands on phase stability are greatly relaxed. 
We solve the problem with a specialized XUV phase-cycling scheme\,\cite{wituschek_tracking_2020}. 
In this scheme, shot-to-shot cycling of the relative carrier-envelope phase between the XUV pulses is implemented which leads to a low-frequency beat note (here $\sim$1\,Hz) in the ion yield. 
Heterodyne lock-in detection is applied using an optical interference signal as the reference waveform for lock-in amplification. 
This detection scheme leads to the downshift of the fringe frequencies by $2 - 4$ orders of magnitude (to $10-100\,$fs) which greatly simplifies the sampling of the interference fringes in the time domain. 
At the same time, the heterodyne detection eliminates most of the phase noise in the signal. 
Thus, it provides a passive interferometer stabilization sufficient for interferometry in the XUV domain. 
As well, background contributions are efficiently suppressed by the lock-in detection technique which greatly improves the sensitivity. 
More details can be found in Refs.\,\cite{tekavec_fluorescence-detected_2007, bruder_phase-modulated_2015, wituschek_tracking_2020}.

\begin{figure}
\centering\includegraphics[width=\linewidth]{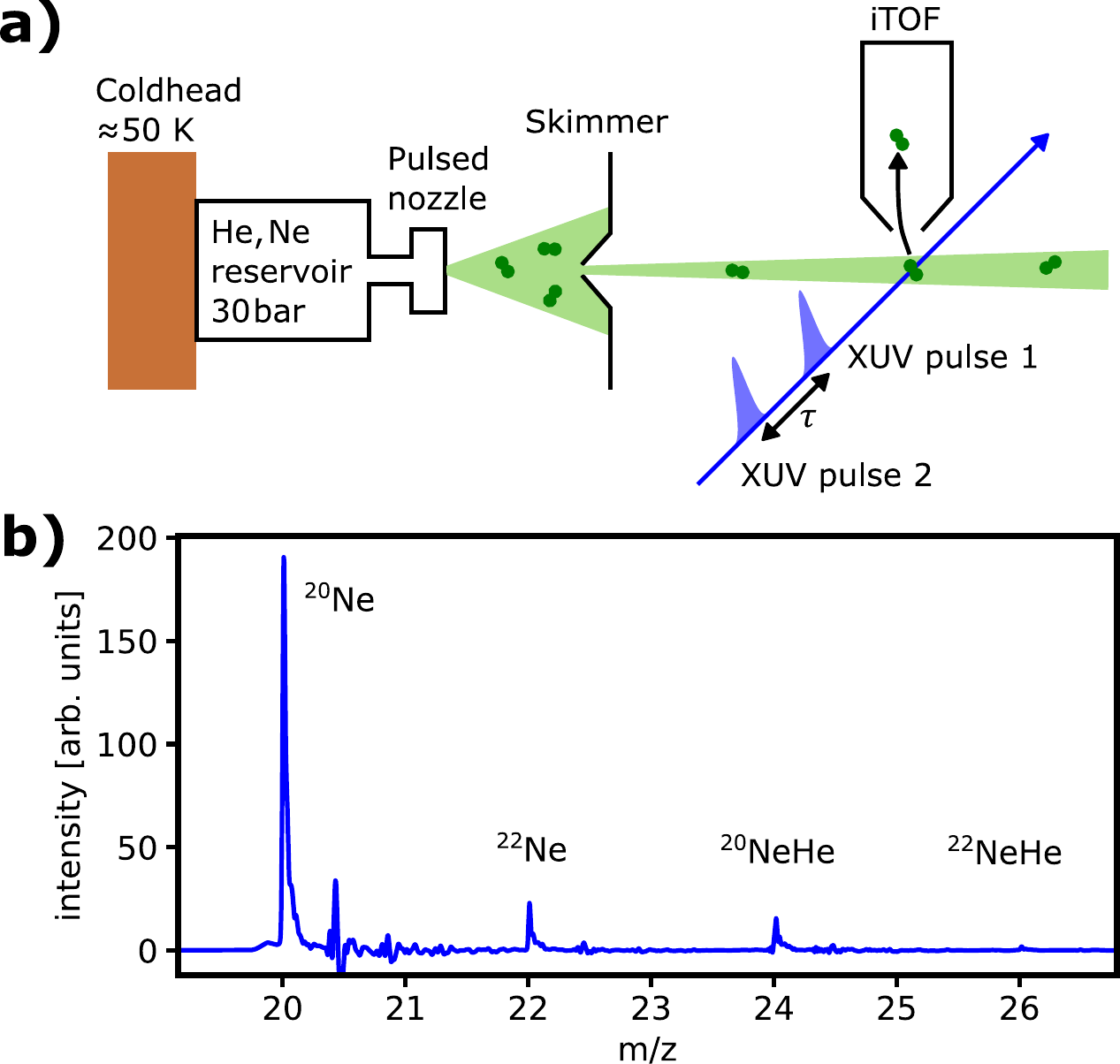}
\caption{(a) Experimental setup. 
(b) Typical ion mass-to-charge (m/z) spectrum for a FEL photon energy of 23.66\,eV. 
Amplification of the detector signal leads to overshoots in the mass spectrum visible, e.g. at m/z=20.5.}
\label{fig3}
\end{figure}
The experiments were performed with the seeded free-electron-laser (FEL) FERMI FEL-1\,\cite{allaria_fermi_2015} at the low-density-matter endstation\,\cite{lyamayev_modular_2013}. 
A compact, highly stable phase-cycling interferometer\,\cite{wituschek_stable_2019} was implemented in the seed-laser beam path for double-pulse seeding\,\cite{gauthier_generation_2016} of the high-gain harmonic generation (HGHG) process. 
In this way, two phase-coherent XUV pulses were generated at the 5th harmonic ($h\nu=23.7\,$eV, $E_\mathrm{pulse}\approx 12\,\mu$J, pulse duration: $\approx$45\,fs, repetition rate: 50\,Hz) whose relative phase was cycled at a rate of 9.3\,Hz. 
No modification of the XUV beam path was necessary and the pulses were automatically collinearly aligned, maximizing the interference contrast. 
The HeNe molecules were produced by co-expansion of a 94\%/6\% mixture of He and Ne gas through a home-built pulsed nozzle (nozzle temperature: 50\,K, orifice diameter: 150\,\textmu m, opening time: 42\,\textmu s) (Fig.\,\ref{fig3}a). 
The FEL pulses intersected with the molecular beam at right angles and the produced ions were recorded with an ion time-of-flight (iTOF) spectrometer.
An exemplary mass spectrum obtained for resonant excitation of the HeNe molecule ($h \nu = 23.66$\,eV) is shown in Fig.\,\ref{fig3}b. 
Under these conditions, the $^{20}$NeHe$^+$ ion yield amounts to  $\approx9$\% of the $^{20}$Ne$^+$ contribution. 
Hence, less than 0.5\% of the species in the gas expansion were HeNe dimers.
For comparison, the pure $^{20}$Ne$_2^+$ ion yield was 19\,\% of the $^{20}$Ne$^+$ yield (not shown). 

The low HeNe particle density and the low signal rates posed particular challenges to the experiment. 
At the used photon energies, we find that only 3\% of the ionized dimers carry an interferometric signal and, thus, show the ICD process. 
In order to extract the interference signals from the mass spectra, the iTOF transients were filtered with a boxcar integrator selecting the He$^{20}$Ne$^+$ mass peak. 
This signal was fed into a lock-in amplifier to selectively amplify only the ion yield exhibiting the characteristic beat note imprinted by the phase cycling of the FEL pulses, while suppressing the background contributions from non-interferometric ionization processes. 
This enabled us to recover the XUV interference signal from the HeNe$^+$ ion yield despite the demanding experimental conditions of the required high phase stability and the dominant background ion yields (see also discussion below). 

\begin{figure}
\centering\includegraphics[width=0.9\linewidth]{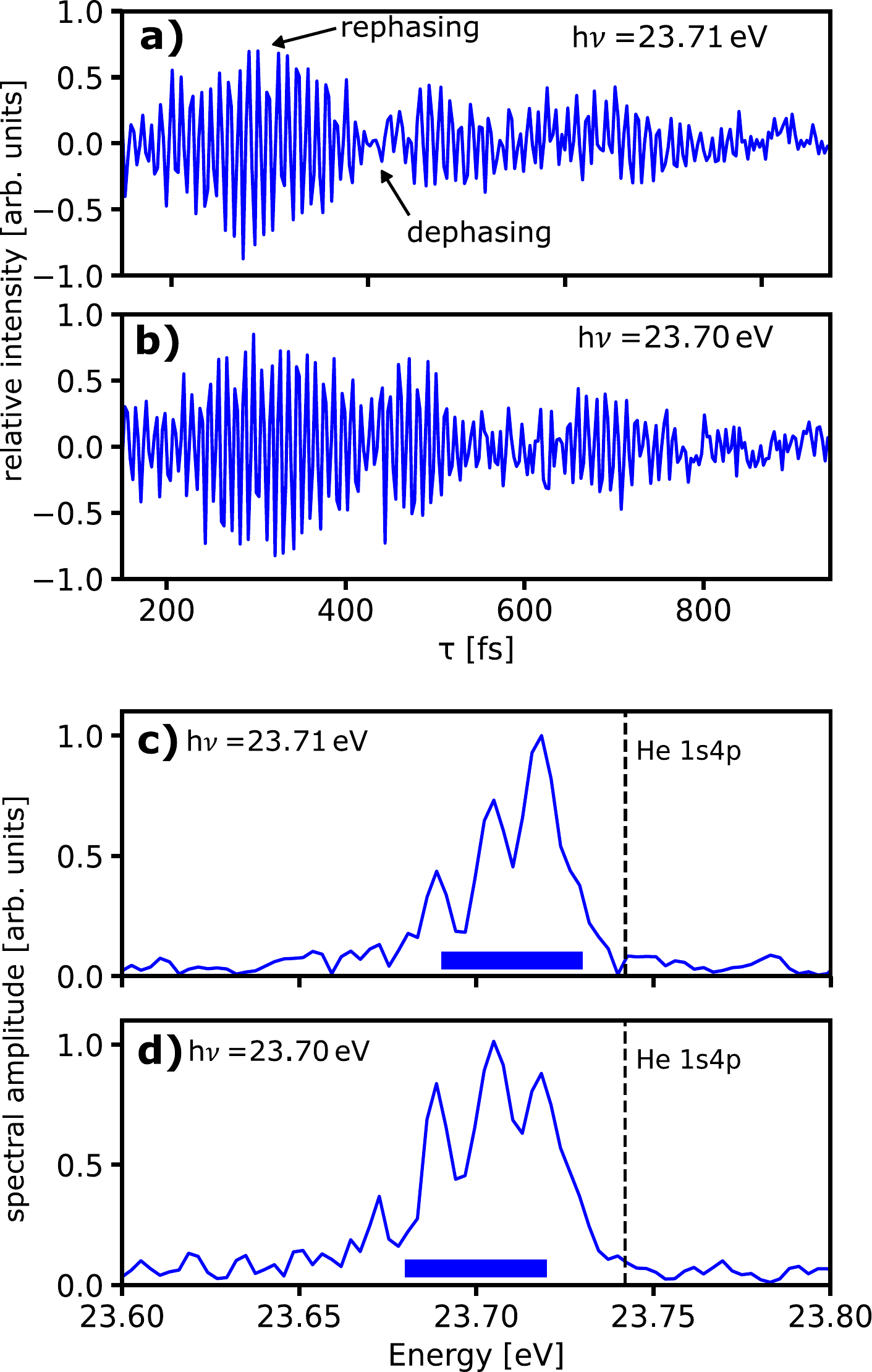}
\caption{(a, b): Interference signals for photon energies $h\nu =23.71$\,eV (a) and 23.70\,eV (b). 
(c, d): Fourier spectra of the interference fringes. 
The FEL center wavelength and spectral bandwidth (FWHM) is indicated by the blue bars, the He $1s \rightarrow 4p$ resonance by the dashed line.}
\label{fig4}
\end{figure}

Figures\,\ref{fig4}\,(a, b) show the interferometric transients for two different excitation energies. 
The interference fringes are clearly visible for inter-pulse delays of up to $\tau = 800$\,fs.
We note that, for $\tau \leq 150$\,fs, optical interference of the seed pulses compromises the HGHG generation and thus the measured interference signal\,\cite{wituschek_high-gain_2020}. 
Therefore, this region is not considered for the data analysis.
The fringe frequency corresponds to the absorption frequency of individual vibronic transitions. 
Due to the heterodyne lock-in detection, the fringe period $\overline{T} \approx 10$\,fs is much larger compared to the fully sampled case, where the fringe period would be $T\approx h/23.7\,eV = 175$\,as. 
Since many different, closely spaced states are excited simultaneously by the broadband XUV pulses, a complex fringe pattern emerges, exhibiting periodic rephasing maxima, which may be interpreted as vibronic WP revivals.
A clear decay of the fringe amplitude can be identified within the observation window which is attributed to the decoherence caused by the ICD process (see below). 
We note, that for the excitation into $n=3$, the ICD process leads to a dissociation of the dimer with a probability of $\approx 30$\,\%\,\cite{trinter_vibrationally_2013, mhamdi_resonant_2018}.  
In the current study, we could not observe any interference signal and thus a signature for the ICD process in the Ne$^+$ yield, probably because of an insufficient signal-to-noise ratio. 

For $n=3$ excitation in HeNe, ICD decay times of $\leq 1$\,ps were deduced\,\cite{trinter_vibrationally_2013} from vibrational line broadenings which is much shorter than the natural lifetime of the excited states ($\sim$\,ns). 
This suggests that also for the excitation to higher-lying asymptotic states of He (with $n >3$), ICD is the dominating decay mechanism causing the decoherence of the interferometric transients in Figs.\,\ref{fig4}\,(a, b). 
Rotational dephasing might be another mechanism explaining the amplitude decay over time. 
In this case, a rephasing of the signal should occur at larger delays beyond the observation window of the current experiment. 
However, the short decay times deduced in the previous study provide a strong indication that the observed decay reflects the decoherence caused by the ICD of the system. 

For the ICD rates a strong dependence on the distance between the interacting constituents ($\propto 1/r^6$)  is expected, leading to a non-exponential decay behavior\,\cite{santra_interatomic_2000,rist_comprehensive_2017}. 
So far, only few time-resolved ICD experiments have been dedicated to this decay behavior\,\cite{fruhling_time-resolved_2015, trinter_ultrafast_2022}.
In principle, the WPI data contains information about the nuclear motion which would allow the real-time mapping of the ICD rate. 
Here, the strong distance dependence of ICD should lead to a step-wise amplitude decay for each round trip of the excited WPs on the potential energy curves (cf. Fig.\,\ref{fig2}b), similar to the observed behavior in dissociating NaI molecules\,\cite{rosker_femtosecond_1988}. 
Accordingly, Figs.\,\ref{fig4}\,(a, b) show a clear amplitude decay on an ultrafast time scale. 
However, for a clear mapping of the nuclear motion and the decay behavior, a disentanglement of the contributions from the different vibrational and electronic states is needed. 
This might be achieved with accurate theory models in combination with an increased signal-to-noise level in the experiment. 
Likewise, a more selective excitation into specific electronic states using narrower XUV pulses may assist this interpretation. 

In principle, conventional femtosecond XUV-pump, NIR-probe photoelectron/-ion spectroscopy should be able to map the pure vibrational WP beatings through the Franck-Condon window to higher-lying cationic states. 
Here, the ICD process should be observable as a loss in the photoionization yield by the probe pulse.
We have attempted these measurements, however, could not discern any vibrational beatings from the background ion yield. 
The large ionization background is explained by the non-resonant ionization of the molecule occurring for the excitation with broadband femtosecond laser pulses at photon energies above the Ne IP ($> 21.56$\,eV), triggering simultaneously the resonant ICD as well as the non-resonant ionization of the molecule. 
This is in contrast to narrow-band synchrotron excitation, where an enhancement of the ionization cross section by a factor of 60 was observed for resonant excitation to states correlated to the He $n=3$ asymptote\,\cite{trinter_vibrationally_2013}. 
In the WPI experiments using broadband femtosecond pulses, the selectivity is greatly enhanced since only interference signals for the resonant excitation contribute outside of the temporal pulse overlap ($\tau \gtrsim 150\,$fs) as shown in Figs.\,\ref{fig4} (a,b). 
Such selectivity is not present in the conventional XUV-pump, NIR-probe experiments. 

As another advantage, by Fourier analysis of the interferometric signal, spectral information beyond the laser bandwidth can be gained. 
Figs.\,\ref{fig4}\,(c, d) show the Fourier transforms of Figs.\,\ref{fig4}\,(a, b), revealing the excitation spectrum of the molecule for both laser wavelengths. 
To recover the absolute energy scale, a linear energy shift of the Fourier spectrum was applied to compensate for the frequency downshift of the heterodyne detection. 
Remarkably, the spectral resolution in the femtosecond interference experiment is 5.5\,meV which is a factor of 7.3 better than the spectral bandwidth of the FEL pulses and only a factor of 3.2 lower than in the high-resolution synchrotron absorption experiment\,\cite{trinter_vibrationally_2013}. 
The high density of electronic and vibrational states in the energy window excited with the femtosecond laser pulses leads to a well-resolved but complex vibronic spectrum. 
In analogy to the synchrotron experiment of the states with $n=3$, calculations of the potential energy curves would provide insight into the individual spectral contributions. 
To the best of our knowledge, the states correlating to the He $n=4$ asymptote have not been calculated for the HeNe dimer and we therefore refrain in the current work from a further assignment of the spectral features. 
We note that, in the applied twin-seeding scheme, an energy chirp on the electron bunch can lead to shifted spectra for the second pulse (up to $\approx 20$\,meV) at large pulse delays as applied here. 
Hence, the amplitudes of the individual features in the Fourier spectrum (Figs.\,\ref{fig4}\,c, d) may be slightly misleading. 
Likewise, the amplitude decay in the time-domain might be influenced by this effect. 
This ambiguity might be reduced by a more careful preparation of the electron bunch and is expected to further improve in echo-enabled harmonic generation\,\cite{rebernik_ribic_coherent_2019}. 

In conclusion, we have used XUV interferometry to probe vibronic WP beatings in autoionizing states of the HeNe molecule on ultrafast time scales. 
The experiment provides a benchmark in sensitivity to probe highly dilute molecular quantum systems with WPI in the XUV domain and reveals the decoherence induced by ultrafast decay processes such as ICD. 
In contrast to narrowband synchrotron radiation, the broadband femtosecond laser excitation launches vibronic WPs extending over several vibrational and electronic states and thus may initiate a concerted motion of the nuclei. 
Combined with the demonstrated Fourier transform analysis, this opens up the possibility to study ICD and related ultrafast processes in real time with high temporal and spectral resolution. 
At the current state of the experiment, a step-wise (rather than smooth exponential) decoherence of the system due to ICD could not be unequivocally identified. 
This would require the precise modeling of the molecule's potential energy curves along with a simulation of the WP propagation and potentially an improvement in the signal-to-noise level of the experiment. 
By adding an NIR ionization pulse in combination with photoelectron detection, the experiment could readily be extended to two-dimensional spectroscopy. 
Such experiments would provide further insights into ultrafast molecular dynamics and would be particularly beneficial for revealing spectral correlations. 
The current study demonstrates a promising step in this direction. 
\section*{Funding}
Funding by the Bundesministerium f\"ur Bildung und Forschung (BMBF) \textit{STAR} (05K19VF3), by the European Research Council (ERC) with the Advanced Grant \textit{COCONIS} (694965) and by the Deutsche Forschungsgemeinschaft (DFG) RTG 2079 and the Research Unit FOR 1789 is acknowledged. Also, funding by the Swedish Research Council, by the Knut and Alice Wallenberg Foundation, Sweden, by the Danish Council for Independent Research (Grant No. 1026-00299B) is acknowledged.

\section*{Acknowledgments}
We gratefully acknowledge the support of the FERMI staff.

\section*{Disclosures}
The authors declare no conflicts of interest.

\bibliography{PM_HeNe_main}

\end{document}